\documentclass[prl,aps,twocolumn,superscriptaddress,showpacs]{revtex4}

\usepackage{epsfig}
\usepackage{amsmath}

\newcommand{\be}{\begin{equation}}
\newcommand{\ee}{\end{equation}}
\newcommand{\bea}{\begin{eqnarray}}
\newcommand{\eea}{\end{eqnarray}}
\newcommand{\la}{\langle}
\newcommand{\ra}{\rangle}
\newcommand{\p}{\partial}
\newcommand{\ham}{{\cal H}}

\begin{document}

\title{Full counting statistics in the self-dual interacting resonant level model}

\author{Sam T. Carr}
\affiliation{\!Institut \!f\"ur Theorie der Kondensierten Materie,\! Karlsruhe Institute of Technology,\! 76128\! Karlsruhe,\! Germany}
\affiliation{DFG Center for Functional Nanostructures, Karlsruhe Institute of Technology, 76128 Karlsruhe, Germany}

\author{Dmitry A. Bagrets}
\affiliation{Institut f\"ur Theoretische Physik, Universit\"at zu K\"oln, 
Z\"ulpicher Str.~77, 50937 K\"oln, Germany}
\affiliation{Institute of Nanotechnology, Karlsruhe Institute of Technology, 76344 Eggenstein-Leopoldshafen, Germany}

\author{Peter Schmitteckert}
\affiliation{DFG Center for Functional Nanostructures, Karlsruhe Institute of Technology, 76128 Karlsruhe, Germany}
\affiliation{Institute of Nanotechnology, Karlsruhe Institute of Technology, 76344 Eggenstein-Leopoldshafen, Germany}

\date{\today}

\pacs{73.63.-b, 72.70.+m, 05.40.Ca, 05.60.Gg} 

\begin{abstract}
We present a general technique to obtain the zero temperature full counting statistics of charge transfer in interacting impurity models out of equilibrium from time-dependent simulations on a lattice. We demonstrate the technique with application to the self-dual interacting resonant level model, where very good agreement between numerical simulations using the density matrix renormalization group and those obtained analytically from the thermodynamic Bethe ansatz is found. We show from the exact form of counting statistics that the quasiparticles involved in transport carry charge $2e$ in the low bias regime, and $e/2$ in the high bias regime. 
\end{abstract}

\maketitle
As nano-devices decrease in size and electron correlations become more important, the study of out-of-equilibrium transport properties 
on the nanoscale becomes of fundamental importance. 
Of great worth are powerful numerical approaches to interacting many-body systems, such as the density matrix renormalization group (DMRG) \cite{DMRG}.
While recent work using the DMRG has shown remarkable development in calculation of $I$-$V$ \cite{Boulat-Saleur-Schmitteckert-2008} and noise \cite{Branschaedel-Boulat-Saleur-Schmitteckert-2010} characteristics of impurity models, 
the grail of such an enterprise is to calculate all moments of charge transfer when the system is driven out of equilibrium, 
known as the full counting statistics (FCS)~\cite{Levitov-Lee-Lesovik-1996, Nazarov-Blanter-book}.

The electron FCS in mesoscopic transport concentrates on a
charge distribution $p_n$, which is the probability that $n$ electrons are transmitted from the left to the right lead during the measurement time $t_m$.   Knowledge of all probabilities $p_n$ gives full knowledge of the statistics of charge transfer.  It is usually more convenient to work with the cumulant generating function (CGF) of the distribution which depends on the {\em counting field} $\chi$ 
and is defined as
\be
Z_{t_m}(\chi) = \sum_n e^{in\chi} p_n = \la e^{i\chi Q/e} \ra.
\ee
The last equality here is well defined in the classical limit, when the charge $Q$ is a random variable associated with 
some stochastic process~\cite{Bagrets-Nazarov-2003}. 
In the quantum case, one has to supplement this definition with a prescription for how the measurements are time-ordered; it was shown in \cite{Levitov-Lee-Lesovik-1996} that in terms of a simple model of a spin-$\frac12$ galvanometer as a measurement device that the correct prescription is
\be
Z_{t_m} (\chi) = \la \tilde{T}_t e^{i \frac{\chi}{2e} \int_0^{t_m} \hat{I}(t) dt }\, {T}_t e^{i \frac{\chi}{2e} \int_0^{t_m} \hat{I}(t) dt } \ra,\label{Zdef}
\ee
where $T_t$ means time-ordered, $\tilde{T}_t$ means anti-time ordered, and the average $\la\ldots\ra$ is taken over some initial (in general non-equilibrium) state.  We deal with $F(\chi)=-\ln Z(\chi)$, as this generates the irreducible cumulants of charge transfer 
$C_n = \left. - \left( \frac{\p}{i\p \chi}\right)^n F(\chi) \right|_{\chi=0}$.
In the long-time limit, each of these is proportional to the measurement time $t_m$, for example $C_1=It_m$ where $I$ is the current flowing in the system, and $C_2=St_m$ where $S$ is the zero-frequency shot noise.

While the FCS for non-interacting electrons is fairly comprehensively understood \cite{Levitov-Lee-Lesovik-1996,Schoenhammer-2007},  far less is known in the presence of interactions \cite{Bagrets-Utsumi-Golubev-Schoen-2006,Gogolin-Komnik-Ludwig-Saleur-2007}, particularly when the interactions drive the system into a strongly correlated regime.  
In this paper we describe a general technique to obtain the CGF for quantum impurity models at finite bias by means of time-dependent numerical simulation.  

As a concrete example, we take the interacting resonant level model (IRLM), described by the Hamiltonian
\begin{gather}
\!\!\!\ham = -t \sum_{n=L,R}  \sum_{i=0}^{L/2} \left( c_{n,i}^\dagger c_{n,i+1} + H. c. \right) + (\epsilon_0 - U)d^\dagger d \label{TheHamiltonian} \\ 
  +  \sum_n \left( t'_n c_{n,0}^\dagger d + H. c \right) + U \sum_n \left( d^\dagger d - \frac{1}{2} \right)
 c_{n,0}^\dagger c_{n,0} . \nonumber
\end{gather}
Here, $c_{n,i}^\dagger$ creates a Fermion on the $i$'th site of the left or right lead, while $d^\dagger$ is the creation operator on the resonant level.  The parameter $t=1$ is the hopping parameter of the leads, $\epsilon_0$ is the energy
of the resonant level, $t'_n$ is the hybridization between the resonant level and the leads 
(here, we will assume a symmetric coupling $t'_R=t'_L$), and $U$ is the interaction between the resonant level and the leads.  In the present work, we concentrate on the point $U=2t$ where the model shows a certain self-duality \cite{Boulat-Saleur-Schmitteckert-2008,Weiss-book}.  The separation of the above Hamiltonian into
 two leads and the quantum impurity 
 however is much more general than this
 model and the method described below may be readily applied to other setups.

The counting field is added via the substitution
$t'_{L(R)} \rightarrow t' e^{\pm i\chi/4}$
and if we define the resulting Hamiltonian after this substitution as $\ham_\chi$, it can be shown that the CGF defined in Eq.~\ref{Zdef} may be rewritten as
\be
Z_{t_m}(\chi) =  \la \Psi(0) | e^{i \ham_{-\chi} t_m} e^{-i \ham_{\chi} t_m} | \Psi(0) \ra.\label{Zcalculate}
\ee
This is now very convenient for a numerical time evolution approach, as $| \Psi(t_m) \ra = e^{-i \ham_{\chi} t_m} | \Psi(0) \ra$ is simply the time evolution of the starting state $\Psi(0)\ra$ -- see also Ref.~\onlinecite{Schoenhammer-2007}. Following  Ref.~\onlinecite{PS:Ann2010},
we prepare an initial state $|\Psi(0) \ra$ at time $\tau=0$ by applying a potential $\pm V_{\mathrm{SD}}/2$ to the left and right leads respectively.  We then switch off the voltage bias and evolve (still with $\chi=0$) until some time $t_0$, when the largest transients have died away, and we are roughly in a steady state.  We then switch on $\chi$ and perform two time evolutions $| \Psi_{\pm \chi}(\tau) \rangle = \exp( -i (\ham_{\pm \chi} - E_0) (\tau - t_0) ) |\Psi(t_0) \rangle$, where $E_0$ is the eigen-energy of the initial state.
  After each time step, we look at the overlap between these two states $\la \Psi_{-\chi}(\tau) | \Psi_{+\chi}(\tau)\ra$, thus producing from Eq.~\ref{Zcalculate} $Z(\chi)$ as a function of the measuring time $t_m=\tau-t_0$.
To obtain the long time  $t_m \rightarrow \infty$ steady state $F(\chi)=t_m \tilde{F}(\chi)$ from the finite time numerical data, we fit to
\be
\frac{\p F(\chi)}{\p t_m}  = \tilde{F}(\chi) + A \cos (V_{\mathrm{SD}} t_m + \alpha) + B e^{-\beta t_m} \cos( \omega t_m + \gamma), 
\ee
where $\frac{\p F(\chi)}{\p t_m}$ is obtained by the  derivative of cubic splines fitted to  $F(\chi)$,
taking care with the complex logarithm so as to get a continuous function.
Here, the first cosine term is the Josephson current that appears from a finite size gap \cite{Boulat-Saleur-Schmitteckert-2008}, 
while the final term is a decaying transient.

\begin{figure}
\begin{center}
\includegraphics[width=3in,clip=true]{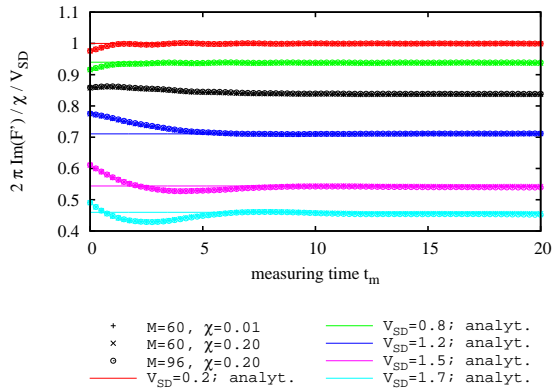}
\end{center}
\caption{[Color online] Time evolution of the imaginary part of $-\p_{t_m} F(\chi) /(\chi V_{\mathrm{SD}})$ for various values of bias voltage $V_{\mathrm{SD}}$, system size $M$, and small values of counting field $\chi$.  As explained in the text, this converges in the long time limit to the conductance of the model.  The analytic result for conductance is also plotted.  It is seen that after an initial transient, excellent agreement is seen between the numerical simulation in the presence of the counting field and the analytic result.}\label{Current}
\end{figure}

We now look at the results of applying this method to the self-dual IRLM, defined by Hamiltonian~(\ref{TheHamiltonian}).  
For the time-evolution, we use the full time-dependent extension \cite{PS:PRB04} of the DMRG   where we optimize the target space for all time steps simultaneously avoiding the run-away errors in the wave function
    in adaptive time evolution schemes \cite{White-2004}.
    The time stepping is performed by a matrix exponential using the Krylov space representation \cite{PS:PRB04}. 
    We first perform short time dynamics up to $t_m=2$ and successively restart the DMRG procedure to longer measuring times up to $t_m=25$ using time steps of $0.25$.
    We keep up to 3100 states per DMRG block leading to target space dimensions of up to $8\cdot 10^6$.
We would like to emphasize that the general framework is not tied to the DMRG and could be applied within other numerical approaches, 
    or even to bosonic problem such as the photon dynamics in wave guiding structures \cite{PSphoton:PRL10}.

\begin{figure}
\begin{center}
\includegraphics[width=3in,clip=true]{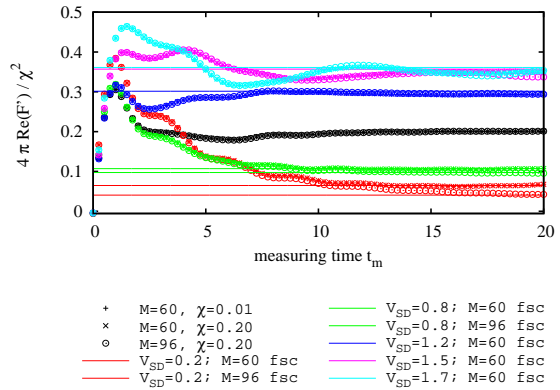}
\end{center}
\caption{[Color online] Time evolution of the real part of $\p_{t_m} 2F(\chi)/\chi^2$ for various values of $V_{\mathrm{SD}}$, $M$ and small values of $\chi$.  
This gives the zero frequency shot-noise of the model, which is known to have a noticeable finite-size correction (fsc) \cite{Branschaedel-Boulat-Saleur-Schmitteckert-2010}. 
Excellent agreement is found between the time evolved simulation and the fsc
 analytic result, which is also plotted.}\label{Noise}
\end{figure}

We begin the discussion by looking at the overlap for small values of $\chi$, where one can write $F(\chi) \approx -i C_1 \chi + C_2 \chi^2/2$.
Hence the imaginary part at $\chi\ll 1$ gives the current.
We plot this in Fig.~\ref{Current} for various values of bias voltage $V_{\mathrm{SD}}$ and counting fields $\chi$, where we see relatively weak transient effects, 
and a clear convergence to the known analytic value, previously discussed in Ref.~\onlinecite{Boulat-Saleur-Schmitteckert-2008}.  
We will return to a discussion of the analytic result later.
Correspondingly, the real part gives shot noise,
which is plotted in Fig.~\ref{Noise}, and can be compared to the results in Ref.~\onlinecite{Branschaedel-Boulat-Saleur-Schmitteckert-2010}.  Despite the rather larger transient behavior, the asymptotic behavior is quite clear, and is in perfect agreement with the previously discussed analytic result for noise (which should be corrected for finite size effects \cite{Branschaedel-Boulat-Saleur-Schmitteckert-2010}). 
We note that within the  present method, the extraction of the (zero-frequency) noise is  a lot easier than obtaining the correlation function directly,
as noise is obtained directly from the large time limit of $\tilde{F}(\chi)$ without having to perform a Fourier transformation including the associated finite size extrapolation \cite{PS:PRB10}.

Having examined a few time scans, we now turn to the FCS -- but before examining the numerical plots let us give the analytical result for the CGF of this model.  It has been long known \cite{Filyov-Wiegmann-1980} that the IRLM is integrable -- that is all eigenvalues and eigenstates of the interacting many-body problem may be constructed via the Bethe ansatz.  However it was only much later realized \cite{Boulat-Saleur-2008} that at the special interaction value given by the self-dual point, the voltage bias operator is also diagonal in this Bethe basis, allowing the exact solution to be found out of equilibrium via the methods of Ref.~\onlinecite{Fendley-Ludwig-Saleur-1995} by mapping to a boundary sine-Gordon (BSG) model.  We note that this is not a generic feature of integrable models, 
a more general extension of the Bethe Ansatz to non-equilibrium systems is currently being discussed in the literature \cite{Andrei}.

The current and shot noise in the self-dual IRLM have previously been found and compared to numerical work in Refs.~\onlinecite{Boulat-Saleur-Schmitteckert-2008,Branschaedel-Boulat-Saleur-Schmitteckert-2010}.  
One can extend the previous analysis~\cite{Saleur-Weiss-2001,Komnik-Saleur-2006}
to obtain all cumulants
of charge transfer in the zero temperature limit.
 We find that the results are best expressed as a recurrence relation \cite{Komnik-Trauzettel-Weiss-2007}
\be
C_{n+1} = \frac{1}{3} \left( C_n - V_{\mathrm{SD}} \frac{\p C_n}{\p V_{\mathrm{SD}}} \right).
\ee
Taking the known expression for the current $C_1$ \cite{Boulat-Saleur-Schmitteckert-2008}  and resumming the cumulants,
we find for small $V_{\mathrm{SD}}<V_c$ :
\begin{subequations}
\bea
\!\!\!\!\tilde{F}(\chi)\!&=&\!
\frac{-i V_{\mathrm{SD}} \chi}{2\pi} \nonumber \\
 \!&-&\!V_{\mathrm{SD}}\!\!\sum_{m > 0}  \frac{a_4(m)}{2m} \left( \frac{V_{\mathrm{SD}}}{T_B'} \right)^{6m} \left( e^{-2mi\chi} -1 \right), \label{Flow} 
\eea
and in the opposite circumstance $V_{\mathrm{SD}}>V_c$:
\be
\tilde{F}(\chi)=
\!V_{\mathrm{SD}}\!\!\sum_{m>0} \frac{2a_{\frac{1}{4}}(m)}{m} \left( \frac{V_{\mathrm{SD}}}{T_B'} \right)^{-\frac{3m}{2}} \!\!\!\left( e^{i\frac{m\chi}{2}}-1 \right). \label{Fhigh}
\ee
\end{subequations}
Here $V_c= \frac{\sqrt{3}}{4^{2/3}}T_B'$ is the convergence radius of each of the series, and $T_B'\sim (t')^{4/3}$ is the natural energy scale of the problem.  The non-universal parameter of proportionality relates the regularization of the field theory (on which the Bethe ansatz solution is based) to that of the original lattice model, and can be taken from previous works \cite{Boulat-Saleur-Schmitteckert-2008}.  This leaves {\em zero free fitting parameters} for all comparisons between analytic and numerical results in the current study. The series coefficients are given by
\be
a_K(m)=\frac{(-1)^{m+1}  \Gamma \left(1+Km \right) }{4\sqrt{\pi}m! \Gamma \left(\frac{3}{2} +(K-1)m\right) }.
\ee
 We note here that the above results could be conjectured from only the self-duality 
 of the model \cite{Weiss-book}.  
We also point out that the formal mapping between the IRLM and the BSG model is only proven at the self-dual
point of the former model, which forces $K=4,1/4$ above.  However a number of other models map
 onto the BSG model over a wider parameter range, to which the present results may be easily adapted.  These include tunneling between quantum-Hall edges \cite{Saleur-Weiss-2001}, impurities in Luttinger liquids \cite{Komnik-Trauzettel-Weiss-2007} and $I$-$V$ characteristics of Josephson junctions \cite{Ingold-Grabert}.

\begin{figure}
\begin{center}
\includegraphics[width=3in,clip=true]{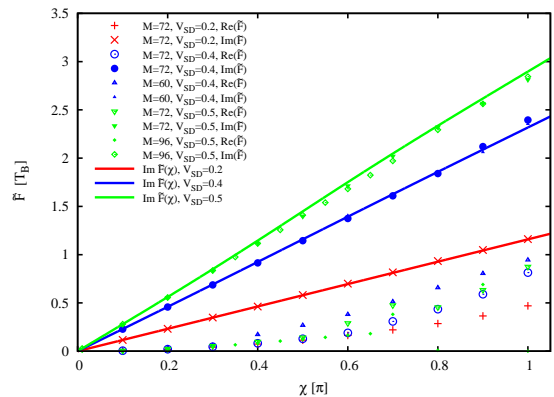}
\end{center}
\caption{[Color online] Cumulant generating function $\tilde{F}(\chi)$ as determined by the DMRG, for various bias voltages $V_{SD}<V_c$.  The data above is for $t'=0.3$ for which $V_c=0.64$.  The corresponding analytic results for the imaginary part is also shown in the plot.  The analytic real part is too small to see on the current scale -- the signal seen by the DMRG is almost entirely due to finite size effects. }\label{IntFh}
\end{figure}

We now analyze the results, first considering the low bias regime.  We plot the numerical data in Fig.~\ref{IntFh} along with the analytic result \eqref{Flow}.  While the numerical data and analytic result show reasonable agreement in the imaginary parts, the real part suffers much more seriously from finite size effects, as was already noticed in Ref.~\onlinecite{Branschaedel-Boulat-Saleur-Schmitteckert-2010}.  In fact, the true values of the real part of $F(\chi)$ in the thermodynamic long time limit are very small, so almost all of the numerical signal is due to the finite size and finite measuring time.  We note that while such large effects might be a nuisance for numerical simulations,  they are also extremely relevant for experiments where transport measurements through nano-devices may exhibit a similar behavior.  
By building a theory of such finite size effects and subtracting this from the data, some preliminary agreement with the analytic result may be seen, in particular the $\pi$ periodicity that will be discussed shortly.  Details of this will be presented elsewhere.

Looking at the analytic expansion, Eq.~\ref{Flow}, we see that the backscattering current may be thought of as a sum of Poissonian processes of effective quasi-particles of charges $2me$ where $m$ is an integer \cite{Saleur-Weiss-2001,Komnik-Trauzettel-Weiss-2007}.  As a result of quantum interference, the effective `probabilities' in this equivalent Poissonian process are not all positive, however from the periodicity of $F(\chi)$, it is clear that the fundamental quasi-particles being scattered carry charge $2e$.   This behavior was already hinted at by the Fano factor \cite{Branschaedel-Boulat-Saleur-Schmitteckert-2010}, knowledge of the full CGF confirms this.

\begin{figure}
\begin{center}
\includegraphics[width=3in,clip=true]{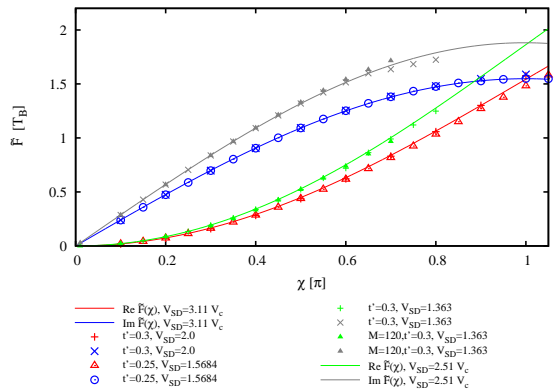}
\end{center}
\caption{[Color online]  Comparison between numerical and analytic results for the cumulant generating function at bias voltages $V_{\mathrm{SD}}>V_c$.  Data for different values of $t'$ such that $V_{\mathrm{SD}}/T_B'$ remains constant collapse onto the same curve, and show excellent agreement with the analytic result.  }\label{IntFl}
\end{figure}

We now turn to the large voltage regime: in Fig.~\ref{IntFl} we plot numerical results along with the appropriate analytic curves from Eq.~\ref{Fhigh}.  Again, we find very good agreement between the two results. As in the low bias voltage case, we can interpret the expansion \eqref{Fhigh} as effective Poissonian processes, but this time with {\em fractionally} charged quasi-particles $me/2$ \cite{Saleur-Weiss-2001,Komnik-Trauzettel-Weiss-2007,Levitov-Reznikov-2004}, again confirming behavior already hinted at by the Fano factor in Ref.~\onlinecite{Branschaedel-Boulat-Saleur-Schmitteckert-2010}.

While the function \eqref{Fhigh} is actually $4\pi$ periodic, it is not clear from the numerical technique of explicitly adding the counting field that we can obtain a result that is not $2\pi$ periodic.  Furthermore, the numerical data is very messy for $\chi>\pi$ making this parameter regime at present unobtainable to us.  As we discuss shortly, the $4\pi$ periodicity formally arises due to  $\tilde{F}(\chi)$ becoming double valued -- the branch may be chosen such that the function is $2\pi$ periodic with discontinuities, or smoothly continued to be $4\pi$ periodic.  In an experimental situation, which branch is measured would depend on details of the setup~\cite{Gutman-Gefen-Mirlin-2010}.
 
Having found effective quasi-particles with charge $e/2$ in the high bias limit, and with charge $2e$ in the low bias limit, we now discuss how to get from one to the other.  In fact, this is easiest to see by studying the counting current $I(\chi)=i\p \tilde{F}/\p \chi$, where we can actually resum the series and obtain the answer in closed form
 \be
 I(\chi) = \frac{V_{\mathrm{SD}}}{2\pi} {}_3F_2 \left[ \{ \frac{1}{4},\frac{3}{4}, 1 \}, \{ \frac{5}{6},\frac{7}{6}\}, - \left( \tilde{V}^3 e^{-i\chi} \right)^2 \right],
 \ee
 where $\tilde V = V_{\mathrm{SD}}/V_c$ is the normalized voltage, and ${}_3F_2$ is a hypergeometric function.  This function has branch cuts along the lines $\chi=\pm \pi/2$ from $\bar{V}=1$ to $\bar{V}=\infty$.  This shows us technically how the CGF transforms from being $\pi$ 
 to $4\pi$ periodic at a {\em critical bias}
  $\tilde{V}=1$.  While the CGF for $|\chi|>\pi/2$ undergoes a sharp transition at this voltage, the 
  curves at low $\chi$, including {\em all  the cumulants}, are smooth and merely show a crossover in their behavior.

In summary, we have presented a numerical technique based on time-dependent DMRG from which one can extract the finite-bias CGF of the FCS for a generic quantum impurity problem.  We have demonstrated the method for the self-dual IRLM, for which we have also obtained results from Bethe ansatz calculations, with good agreement.  From this, we provide a scenario of obtaining charge fractionalization in transport experiments, which we relate to the analytic properties of the CGF.

The numerics was performed at the XC2 and HC3 of the Steinbuch Centre of Computing, Karlsruhe,
within project RT-DMRG.

\vspace{-0.5cm}


\end{document}